\documentclass[preprint,12pt]{elsarticle}

\usepackage{amssymb,float}
\usepackage{amsmath,amssymb,graphicx}

\journal{}

\begin{document}

\begin{frontmatter}



\title{Performance comparison of computational ghost imaging versus single-pixel camera in light disturbance environment}


\author[label1]{Wenlin Gong \corref{cor1}}

\address[label1]{School of Optoelectronic Science and Engineering, Soochow University, Suzhou 215006, China}
\cortext[cor1]{Corresponding author: wlgong@suda.edu.cn, gongwl@siom.ac.cn.}
\begin{abstract}
Computational ghost imaging (CGI) and single-pixel camera (SPC) are two types of single-pixel imaging and attracts increasing interest in recent years. The performance differences of CGI and SPC in difference light disturbance environment are investigated. In comparison with CGI, we demonstrate that the quality of SPC is always better in the same conditions and local light disturbance has a larger influence to CGI/SPC than global light disturbance. In addition, a method to improve the reconstruction result of SPC is proposed if the source's energy is unstable and there is light disturbance in the detection process, and its validity is also verified by numerical simulation.
\end{abstract}



\begin{keyword}
Computational ghost imaging \sep Single-pixel camera \sep Image reconstruction \sep Light disturbance



\end{keyword}

\end{frontmatter}

\section{Introduction}
\label{}
Single-pixel imaging is a technology that produces images by computing the correlation function between the intensity of modulation field and the target's transmitted/reflected intensity recorded by a detector without spatial resolution \cite{Cheng,Bennink,Cao,Angelo,Shapiro,Graham-Rowe,Duarte,Edgar}. Computational ghost imaging (CGI) and single-pixel camera (SPC), as two types of single-pixel imaging method, have attracted lots of attentions in recent two decades \cite{Shapiro,Edgar}. The difference between CGI and SPC in the optical structure is the position of the modulation device. For CGI, the modulation light field is before the target whereas the target's image is imaged onto the modulation device for SPC \cite{Bennink,Graham-Rowe,Edgar}. Up to now, the feasibility of CGI/SPC has been experimentally demonstrated from X-rays to microwave sources \cite{Zhao,Erkmen,Wang,Sun,Gong1,Sun1,Yu,Pelliccia,Zhang,Liu,Chan,Wang1}, and great development has been achieved especially in the area of remote sensing \cite{Zhao,Erkmen,Wang}, three-dimensional imaging \cite{Sun,Gong1,Sun1}, X-ray imaging \cite{Yu,Pelliccia,Zhang}, and optical encryption \cite{Clemente}.

In essence, both CGI and SPC express the spatial distribution of an object's function as a linear combination of basis patterns \cite{Gong2,Wang2,Candes}. When the imaging system has no disturbance and all the photons transmitted/reflected from the target can be collected by the bucket detector, CGI is equivalent to SPC \cite{Edgar,Sun2}. However, in practical applications, on the one hand we can only receive part of photons from the target because of the limitation of the receiving system's optical aperture especially in long-range detection \cite{Zhao,Gong1,Wang1}. On the other hand, the source's energy fluctuation and the light disturbance from the environment is inevitable. Therefore, the imaging property of CGI and SPC will be different and it is important to clarify their disadvantages and advantages for applications. According to the optical structures of CGI and SPC, SPC has some obvious advantages in comparison with CGI. For example, SPC usually satisfies the process of bucket detection in long-range imaging, and for CGI the energy of the source is limited by the damage threshold of the modulator so that the detection range is restricted \cite{Sun2,Mei}. Recently, some works on the influence of the source's energy fluctuation and the background light to CGI have been reported \cite{Mei1,Deng,Zhou,Li,Yang}, but there is no investigation on the performance comparison between SPC and CGI. In this paper, we study the influence of some different types of light disturbance to both CGI and SPC, and the method to further enhance the quality of SPC is also proposed. What's more, the performance differences between CGI and SPC in light disturbance environment are validated by numerical simulation.

\section{Model and reconstruction}

\begin{figure}[H]
\centering
\includegraphics[width=12.0cm]{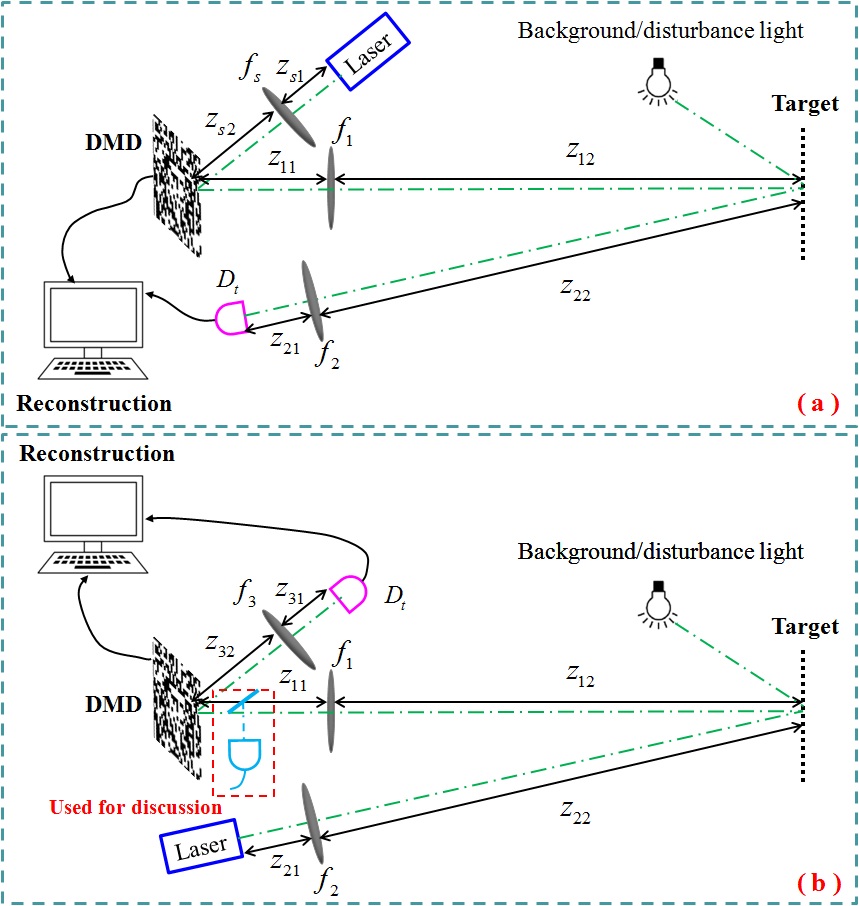}
\caption{Schematics of computational ghost imaging (a) and single-pixel camera (b) in background or disturbance light environment.}
\end{figure}

For CGI in light disturbance environment, as shown in Fig. 1(a), the light emitted from a pulsed laser uniformly illuminates a digital micro-mirror device (DMD) and a series of random coded patterns are prebuilt by modulating the mirrors of the DMD. Then the patterns reflected by the DMD are imaged onto a target by an optical imaging system with the focal length $f_1$, and the photons reflected from the target are collected onto a bucket detector $D_t$ by using another conventional imaging system with the focal length $f_2$. Furthermore, a random light field simultaneously illuminates the same target to analysis the influence of the light disturbance to CGI. The intensity $B_{\rm{CGI}}^i$ recorded by the detector $D_t$ can be represented as \cite{Goodman}
\begin{eqnarray}
B_{\rm{CGI}}^i  &=& \int {\left( {I_{\rm{CGI}}^i (x)+I_b^i (x)} \right)T(x)} dx, {\rm{ \ }} \forall _i  = 1 \cdots K,
\end{eqnarray}
where $T(x)$ denotes the intensity reflection function of the object and $I_{\rm{CGI}}^i (x)$ is the intensity distribution of the speckle pattern modulated by DMD for the $i$th measurement. In addition, $I_b^i (x)$ is the intensity distribution of the disturbance light field illuminating on the object plane and $K$ is the total measurement number.

SPC in light disturbance environment is displayed in Fig. 1(b), where the positions of both the bucket detector $D_t$ and the pulsed laser are exchanged in comparison with Fig. 1(a). The detection progress is described as

\begin{eqnarray}
B_{\rm{SPC}}^i  &=& \int {I_{\rm{SPC}}^i (x)\left( {I_0+I_b^i (x)} \right)T(x)} dx, {\rm{ \ }} \forall _i  = 1 \cdots K,
\end{eqnarray}
where $I_{\rm{SPC}}^i (x)$ denotes the intensity distribution of the speckle pattern modulated by DMD for the $i$th measurement and $I_0$ is the intensity of the pulsed laser illuminating on the object plane.

According to the principle of single-pixel imaging, the object's image $O_{\rm{GI/SPC}}$ can be reconstructed by computing the correlation function between the pattern's intensity distributions $I_{\rm{CGI/SPC}}^i(x)$ modulated by the DMD and the intensities $B_{\rm{CGI/SPC}}^i$ recorded by the detector $D_t$ \cite{Cheng,Angelo,Gong2}

\begin{eqnarray}
O_{\rm{CGI/SPC}}(x) &=& \frac{1}{{K }}\sum\limits_{i =1}^{K}\left( {I_{\rm{CGI/SPC}}^i(x)-\left\langle {I_{\rm{CGI/SPC}}(x)} \right\rangle} \right)B_{\rm{CGI/SPC}}^i.
\end{eqnarray}
where ${\left\langle {I_{\rm{GI/SPC}}(x)} \right\rangle} =\frac{1}{{K }}\sum\limits_{s = 1}^{K}I_{\rm{GI/SPC}}^i(x)$ represents the ensemble average of $I_{\rm{GI/SPC}}^i(x)$.

In order to evaluate quantitatively the quality of images reconstructed by CGI and SPC, the reconstruction fidelity is estimated by calculating the peak signal-to-noise ratio (PSNR):
\begin{eqnarray}
{\rm{PSNR} } = 10 \times \log _{10} \left[ {\frac{{(2^p  - 1)^2 }}{{{\rm{MSE} }}}} \right].
\end{eqnarray}
where the bigger the value PSNR is, the better the quality of the recovered image is. For a 0$\sim$255 gray-scale image, $p$=8 and MSE represents the mean square error of the reconstruction images $O_{\rm{rec}}$ with respect to the original object $O$, namely
\begin{eqnarray}
{\rm{MSE} }=\frac{1}{{N_{pix}}}\sum\limits_{i = 1}^{N_{pix}}{\left[ {O_{{\rm{rec }}} (x_i) - O (x_i)} \right]} ^2.
\end{eqnarray}
where $N_{pix}$ is the total pixel number of the image.

\section{Simulation demonstration and discussion}

To demonstrate the performance differences between CGI and SPC in the case of light disturbance environment, the parameters of numerical simulation based on the schematic of Fig. 1 are set as follows: the wavelength of the laser is 532 nm, the transverse size of the patterns at the DMD plane is set as 54.6 $\mu$m and the modulated area of the DMD is 64$\times$64 pixels (one pixel is equal to the pattern's transverse size). The speckles modulated by the DMD are hadamard patterns and the measurement number K=4096. In addition, $z_{11}$=$z_{12}$=$z_{21}$=$z_{22}$=200 mm, $f_1$=$f_2$=100 mm. The imaging target, as illustrated in Fig. 2, is a gray-scale object (``siom", 64$\times$64 pixels). According to Eq. (1) and Eq. (2), the irradiation SNR $\varepsilon  = \frac{{\left\langle {I_{\rm{CGI}}^i(x) } \right\rangle }}{{\left\langle {I_b^i(x) } \right\rangle }}$ denotes the signal power to light disturbance power ratio. When the light disturbance obeys spatially uniform distribution in statistics and its transverse size is also 54.6 $\mu$m, Fig. 2 and Fig. 3 have given the influence of spatial global and local light disturbance to both CGI and SPC in different irradiation SNR $\varepsilon$, respectively. Here global light disturbance means that the field of view (FOV) of the disturbance light is larger than the target's area (see the pink rectangle area labeled in Fig. 2), whereas the FOV of the disturbance light is far smaller than the target's area for local light disturbance (see the pink rectangle area labeled in Fig. 3). It is clearly seen that the quality of SPC is always better than that of CGI in the same $\varepsilon$, and local light disturbance has a larger influence to SPC/CGI compared with global light disturbance. For example, as shown in Fig. 2(a), Fig. 2(d) and Fig. 2(g), the reconstruction result of SPC in the case of $\varepsilon$=-20 dB is better than that of CGI in the case of $\varepsilon$=0 dB. And for local light disturbance, SPC is disabled in the condition of $\varepsilon$=-15 dB (Fig. 3(a)) whereas the reconstruction PSNR of SPC can reach to 18.46 dB (Fig. 2(a) and Fig. 2(g)) for global light disturbance even if $\varepsilon$ is -20 dB.

\begin{figure}[H]
\centering\includegraphics[width=14cm]{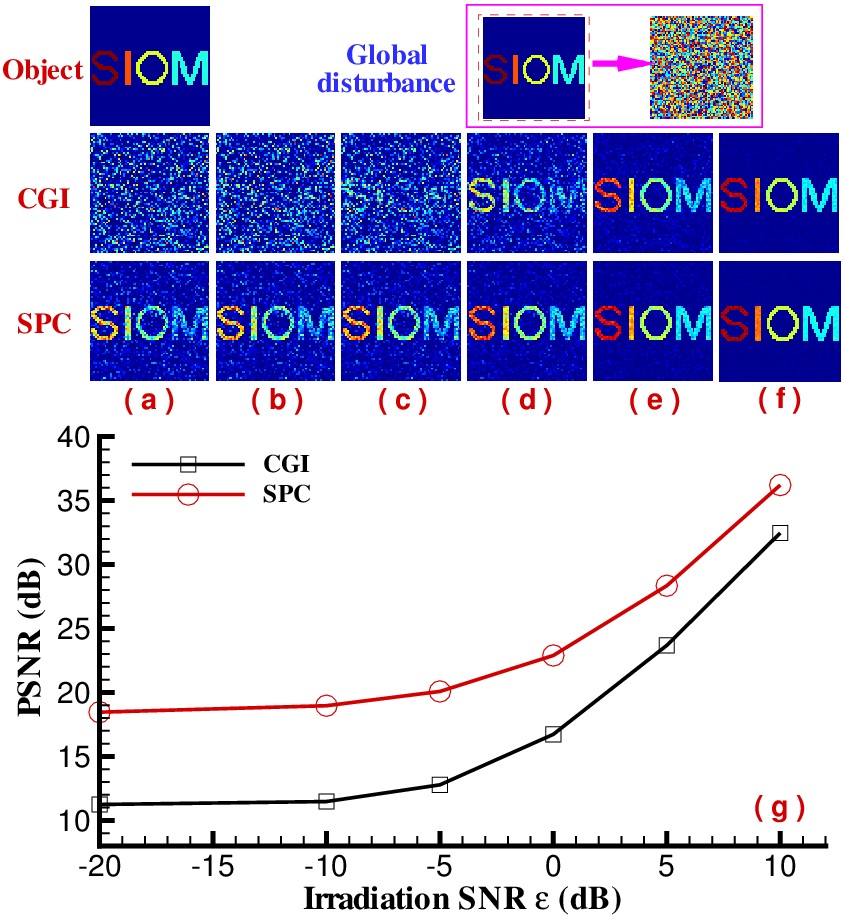}
\caption{The effect of spatial global light disturbance on CGI and SPC in different irradiation SNR $\varepsilon$. (a) $\varepsilon$=-20 dB; (b) $\varepsilon$=-10 dB; (c) $\varepsilon$=-5 dB; (d) $\varepsilon$=0 dB; (e) $\varepsilon$=5 dB; (f) $\varepsilon$=10 dB; (g) the curve of PSNR-$\varepsilon$.}
\end{figure}

\begin{figure}[H]
\centering\includegraphics[width=12cm]{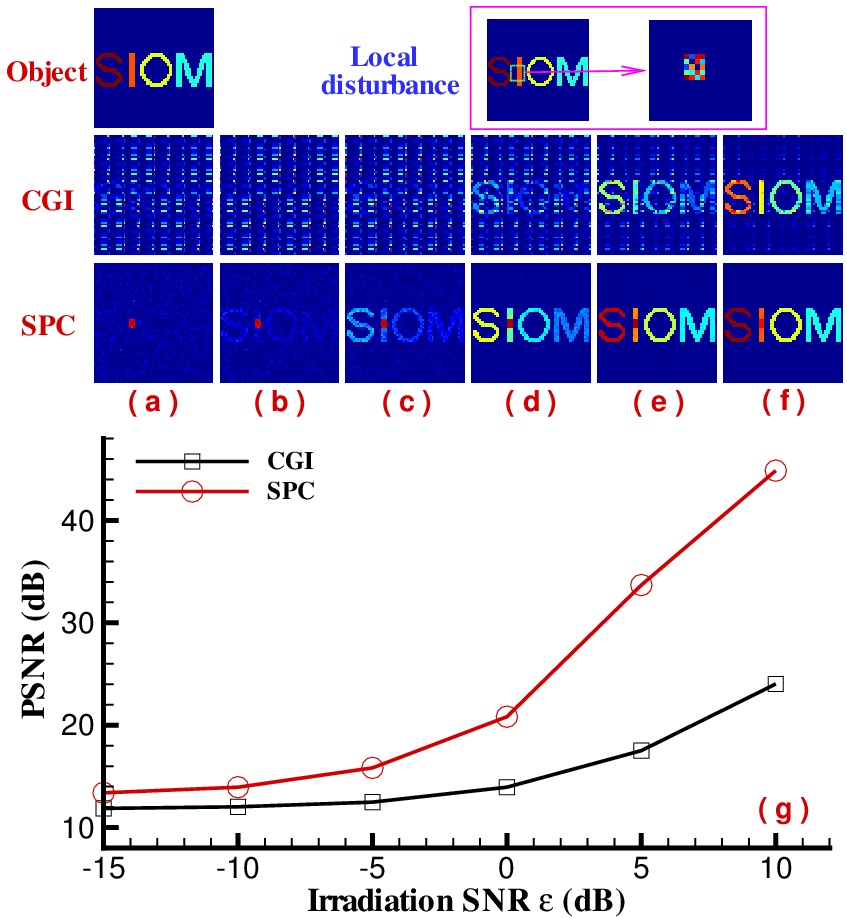}
\caption{The influence of spatial local light disturbance to CGI and SPC in different irradiation SNR $\varepsilon$. (a) $\varepsilon$=-15 dB; (b) $\varepsilon$=-10 dB; (c) $\varepsilon$=-5 dB; (d) $\varepsilon$=0 dB; (e) $\varepsilon$=5 dB; (f) $\varepsilon$=10 dB; (g) the curve of PSNR-$\varepsilon$.}
\end{figure}

In Fig. 4, the light disturbance is considered to be global and uniform for each illumination, but its intensity obeys normal distribution. The intensity disturbance degree of the global light disturbance is defined as $\gamma=\frac{{std\left( {I_b^i(x)} \right) }}{{I_0 }}$ (where $std\left( {I_b^i(x)} \right)$ denotes the standard deviation of $I_b^i(x)$). Using the same gray-scale object shown in Fig. 2 as the imaging target, the reconstruction results of CGI/SPC in different intensity disturbance degree $\gamma$ are illustrated in Fig. 4(a)-Fig. 4(f) and the curve of PSNR-$\gamma$ is displayed in Fig. 4(g). It is obviously observed that SPC is also superior to CGI, which is similar to the results illustrated in Fig. 2 and Fig. 3.

\begin{figure}[H]
\centering\includegraphics[width=12cm]{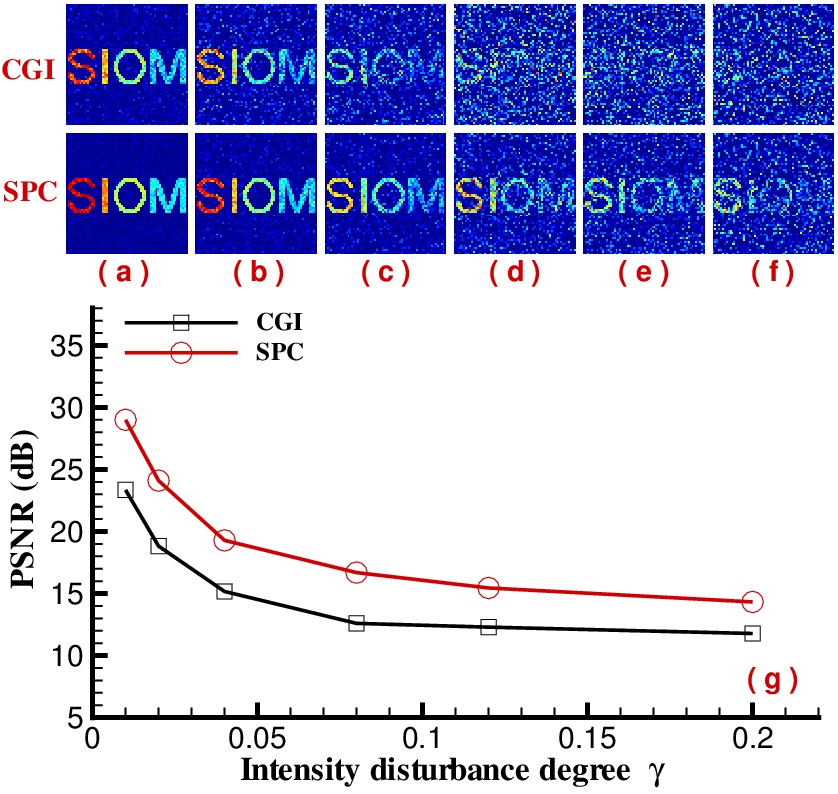}
\caption{Simulated demonstration results of the relationship between the intensity disturbance degree $\gamma$ and the reconstruction results of CGI/SPC. (a) $\gamma$=0.01; (b) $\gamma$=0.02; (c) $\gamma$=0.04; (d) $\gamma$=0.08; (e) $\gamma$=0.12; (f) $\gamma$=0.2; (g) the curve of PSNR-$\gamma$.}
\end{figure}

For SPC and CGI, as displayed in Eqs. (1)-(3), the source's energy fluctuation and the light disturbance will lead to the degradation of imaging quality. For CGI, the light disturbance can be considered as a random additive noise while the source's energy fluctuation can be regarded as a random multiplicative noise. As described in Ref. \cite{Mei1}, the source's energy fluctuation can be measured by a monitor and its influence to CGI can be overcome by correcting the detection signal $B_{\rm{CGI}}^i$ in the process of image reconstruction. However, the effect of the light disturbance on CGI is hard to solve because the photons reflected from the disturbance light and the signal light are mixed. Based on the idea of CGI described in Ref. \cite{Mei1} and the property of SPC, as shown in Fig. 1(b), if a small part of the energy reflected from the target is divided by a beam splitter and is detected by a monitor before modulation, the reconstruction quality of SPC may be improved when the intensity $B_{\rm{SPC}}^i$ is corrected by the monitored signal in the image reconstruction process. Fig. 5 has illustrated the results of SPC and $\rm{SPC}_{\rm{correction}}$ in difference light disturbance cases. As displayed in Fig. 5(a) and Fig. 5(b), the quality of SPC has a little enhancement by the method. However, $\rm{SPC}_{\rm{correction}}$ can be dramatically improved compared with SPC if there is only intensity fluctuation for the light disturbance (Fig. 5(c)). The reason is that the intensity fluctuation of the term $I_0+I_b^i (x)$ in Eq. (2) for each measurement is measured for SPC and it is removed in the process of image reconstruction. Different from CGI described in Ref. \cite{Mei1}, the energy fluctuation of both the source and the disturbance light is monitored by the method of $\rm{SPC}_{\rm{correction}}$, so $\rm{SPC}_{\rm{correction}}$ may be also valid to high-resolution imaging in the case of atmospheric scintillation.

\begin{figure}[H]
\centering\includegraphics[width=14cm]{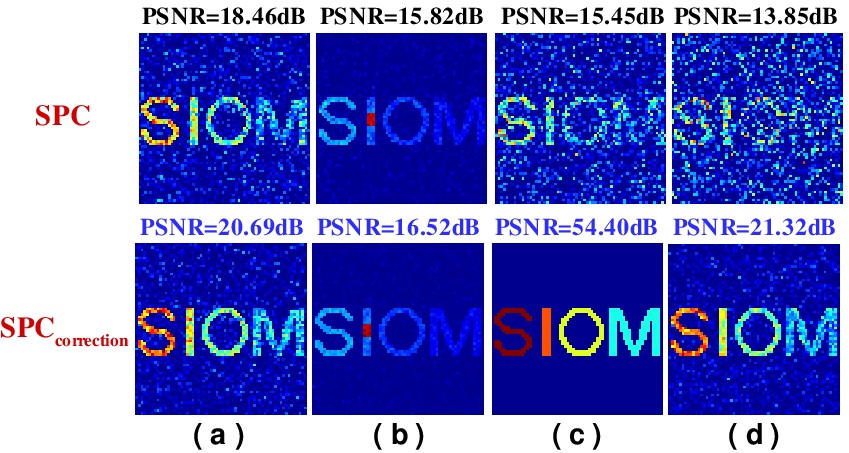}
\caption{Performance comparison of SPC and $\rm{SPC}_{\rm{correction}}$ in difference light disturbance cases. (a) Spatial global light disturbance, $\varepsilon$=-20 dB; (b) spatial local light disturbance, $\varepsilon$=-5 dB; (c) global intensity disturbance, $\gamma$=0.12; (d) global spatial and intensity light disturbance, $\varepsilon$=-10 dB and $\gamma$=0.2.}
\end{figure}

\section{Conclusion}

In conclusion, we have clarified that the types of light disturbance have a different effect on CGI/SPC and shown that SPC is always superior to CGI in light disturbance environment. When a monitor is used to measure the intensity fluctuation of the target's reflection signal before modulation and the single-pixel signal of SPC is corrected in the process of image reconstruction, the quality of SPC can be further enhanced. The work is helpful to the system design of single-pixel imaging in practical applications.

\section*{Funding}

This work is supported by Natural Science Research Project of Colleges and Universities in Jiangsu Province (21KJA140001), Aeronautical Science Foundation of China (2020Z073012001), and Startup Funding of Soochow University (NH15900221).

\end{document}